\documentclass[conference]{IEEEtran}
\IEEEoverridecommandlockouts
\bstctlcite{IEEEexample:BSTcontrol}
\usepackage{cite}
\usepackage{amsmath,amssymb,amsfonts}
\usepackage{algorithmic}
\usepackage{graphicx}
\usepackage{textcomp}
\usepackage{float}
\usepackage{booktabs}
\usepackage{tabularx}
\usepackage{etoolbox}
\def\BibTeX{{\rm B\kern-.05em{\sc i\kern-.025em b}\kern-.08em
    T\kern-.1667em\lower.7ex\hbox{E}\kern-.125emX}}

\usepackage{eso-pic} 

\begin{document}

\AddToShipoutPictureBG*{
\AtPageUpperLeft{
\setlength\unitlength{1in}
\hspace*{\dimexpr0.5\paperwidth\relax}
\makebox(0,-0.75)[c]{\textbf{
}}}} 

\title{Analyzing behavioral trends in community driven discussion platforms like Reddit}

\author{\IEEEauthorblockN{Sachin Thukral}
\IEEEauthorblockA{\textit{TCS Research} \\
New Delhi, India \\
sachi.2@tcs.com}\and
\IEEEauthorblockN{Hardik Meisheri}
\IEEEauthorblockA{\textit{TCS Research} \\
New Delhi, India \\
hardik.meisheri@tcs.com}\\
\IEEEauthorblockN{Arnab Chatterjee}
\IEEEauthorblockA{\textit{TCS Research} \\
New Delhi, India \\
arnab.chatterjee4@tcs.com}
\and
\IEEEauthorblockN{Tushar Kataria}
\IEEEauthorblockA{\textit{IIIT Delhi} \\
New Delhi, India \\
tushar15184@iiitd.ac.in}
\and
\IEEEauthorblockN{Aman Agarwal}
\IEEEauthorblockA{\textit{IIIT Delhi} \\
New Delhi, India \\
aman15012@iiitd.ac.in}\\
\IEEEauthorblockN{Lipika Dey}
\IEEEauthorblockA{\textit{TCS Research} \\
New Delhi, India \\
lipika.dey@tcs.com}
\and

\IEEEauthorblockN{Ishan Verma}
\IEEEauthorblockA{\textit{TCS Research} \\
New Delhi, India \\
ishan.verma@tcs.com}

}

\maketitle
 
\begin{abstract}

The aim of this paper is to present methods to systematically analyze individual and group behavioral patterns observed in community driven discussion platforms like Reddit where users exchange information and views on various topics of current interest. We conduct this study by analyzing the statistical behavior of posts and modeling user interactions around them. We have chosen Reddit as an example, since it has grown exponentially from a small community to one of the biggest social network platforms in the recent times. Due to its large user base and popularity, a variety of behavior is present among users in terms of their activity. Our study provides interesting insights about a large number of inactive posts which fail to gather attention despite their authors exhibiting \textit{Cyborg-like} behavior to draw attention. We also present interesting insights about short-lived but extremely active posts emulating a phenomenon like \textit{Mayfly Buzz}. Further, we present methods to find the nature of activity around highly active posts to determine the presence of \textit{Limelight} hogging activity, if any. We analyzed over $2$ million posts and more than $7$ million user responses to them during entire 2008 and over $63$ million posts and over $608$ million user responses to them from August 2014 to July 2015 amounting to two one-year periods, in order to understand how social media space has evolved over the years. 

\end{abstract}


\section{Introduction}

The analysis of human social interactions data has gained much attention in the last two decades, primarily due to the availability of massive amounts of data from the electronic footprints of human social behavior on a variety of online social platforms and the important insights gained from multidisciplinary approaches as well. Complex network analysis~\cite{albert2002statistical,newman2011structure} merging with the traditional approaches in social sciences, along with tools and formalisms from a variety of disciplines like theoretical physics, applied mathematics, statistics and even psychology with a core of computer science, have developed into what is currently popular as computational social science~\cite{lazer2009computational}. The current approach to social network analysis has much built upon the classical approaches~\cite{wasserman1994social}, and the present interests in online social networks span across operations research, market intelligence, survey science, and statistical computing.

Study of social network data not only reveal the structure of the connected components including strong and weak ties and their dynamics, but also the possible reasons as to why such structure and dynamics are prevalent.


Any social network can be seen as a multi-dimensional graph where elements like posts, comments, users etc. act as nodes and their interaction form the links. Statistics like post lifespan, average number of posts per unit time shows an aggregate user behavior along the post dimension across the entire social media platform. User comments across posts give it an interactivity flavor where user behavior can be segregated according to number of comments to time span of interaction. Adding a layer of the number of distinct users involved opens the scope to differentiate user behavior in terms of reachability and impact of the post. 

If one has access to a huge amount of data, a rigorous statistical analysis combined with  behavioral studies can bring out interesting features from the data, both spatially and temporally, providing interesting insights. In this paper, we are:
\begin{enumerate}

\item Studying evolution patterns of posts over time based on user interactions with the posts and grouping them into different categories,

\item Categorizing posts based on user interaction patterns emerging around them. We present methods to determine the focal points of interactions,  

\item Presenting methods to identify behavioral trends exhibited by users in order to popularize their posts.

\end{enumerate}

In the social news aggregation, web content rating and discussion website \textit{Reddit.com}, members share content in the form of links, text posts and images, which are then voted up or down by other members, wherefrom further discussions can emerge. Posts cover a variety of topics including news, science, movies, video games, music, books, fitness, food, and image-sharing. They are organized by subject into user-created \textit{subreddits}, providing further opportunities for fostering discussion, raising attention and publicity for causes.
While Reddit is known for its open nature and diverse user community across different  demographics and subcultures that generate its content, posts are also moderated for various reasons.

In this paper, we have tried to gather insights about where, when and by whom the content is being driven in the community as a whole. Studying evolution patterns help us in understanding characteristics of posts garnering huge number of responses. Studying characteristics from a authors perspective gives us an indication of which authors are more reliable in spreading information over the space. On the similar lines, identifying focal points in a long discussion can lead to understand popular opinions. These markers and behavioral trends can be used as cues in various applications like advertisement placement, summarizing viral/popular topics from different perspectives, half life of information spread, etc. 

With the increasing use of social media for collaboration and sharing of important information even within  enterprises, understanding human behaviors and able to characterize them as well as understand the interaction dynamics among a group of users is itself turning out to be an important task. 
For example, the organization to which most of the authors of this work are affiliated to, more than $400,000$ employees engage in at least two organization specific, closed social networks serving different purposes. Analysis of temporal patterns and group dynamics presented in our work are important aspects which can not only aid in understanding the different categories of users, but also identify the information needs and push the right content or advertisement for the right group at the right time. The similarity of patterns observed over multiple data sources prove that user behaviors are fairly similar across social platforms.


The rest of the paper is organized as follows: Section~\ref{sec:related} presents the earlier related work. A brief description of Reddit data used in our study has been presented in Section~\ref{sec:data}. Section~\ref{sec:aggr} present aggregate analysis of the data, which provides the basis for our further analysis. Analysis of evolution patterns of posts is presented in Section~\ref{sec:evolution_patterns}. Section~\ref{sec:interaction} shows the interaction dynamics, while Section~\ref{sec:author} shows the behavior exhibited by authors over the space. 
Finally, a summary of the entire analysis and the inferences drawn are presented in   Section~\ref{sec:conclusion}.



\section{Related Work}
\label{sec:related}
There have been many studies regarding social media dynamics from various perspectives. In one of such studies, authors have examined the structure of the comment threads by analyzing the radial tree representation of thread hierarchies~\cite{gomez2008statistical}, while another study presented the responses over a post using graph theoretic approach to infer the \textit{for} and \textit{against} communities for that particular post~\cite{agrawal2003mining}. The basic assumption made for the study was that every post contains at least one comment belonging to each community. 

Researchers have studied the behavioral aspects of users by crowd-sourcing information from experiments on the platform. One such study focuses on how individuals consume information through social news websites and contribute to their ranking systems. A study on the browsing and rating pattern reported that most users do not read the article that they vote on, and in fact 73\% of posts were rated without first viewing the content~\cite{glenski2017consumers}.  While user interactions (likes, votes, clicks, and views) serve as a proxy for the content's quality, popularity, or news-worthiness, predicting user behavior was found to be quite easy~\cite{glenski2017predicting}. A study on the voting pattern in the Reddit~\cite{mills2011researching} has been studied to analyse the upranking of posts from the new page to front page and behavior of users towards some posts which are getting positive or negative votings. They have studied the posts mentioning Wikileaks and Fox News and see the impact of negative voting on them, although analyzing only one month of data. A study on rating effect on posts and comments~\cite{glenski2017rating} has revealed that random rating manipulations on posts and comments led to significant changes in downstream ratings leading to significantly different final outcomes -- positive herding effects for positive treatments on posts, increasing final ratings on the average, but not for positive treatments on comments, while negative herding effects for negative treatments on posts and comments, decreasing the final ratings on average. An exploratory study~\cite{weninger2013exploration} on the dynamics of discussion threads found topical hierarchy in discussion threads, and their possibility to be used to enhance Web search. A study on `social roles' of users~\cite{buntain2014identifying} found that the typical ``answer person'' role is quite prominent, while such individual users are not active beyond one particular subreddit.

In one study, authors have used the volume of comments a blog post receives as a notion of popularity to model the relationship with the text~\cite{Worthy_of_comment}. Authors used various regression models to predict the volume of comments given in the text. This analysis is restricted in terms of dataset scale, limited to political posts and three sites which amount to four thousand posts. While content analysis is most intuitive, it does not provide richer analysis. Text content shared over social media is noisy, full of non-standard grammar and spelling, often cryptic and uninformative to the outsider from the community. When one adds the scale of today's social media dataset it is computationally non-viable to have content analysis over the whole corpus.

Most of the studies reported till date have performed analysis on a subset of data by restricting themselves to a limited number of posts, comments, top users, subreddits etc., while we use the complete data for an entire one year period. To the best of our knowledge, only, very few have used complete data for analysis. In Ref.~\cite{singer2014evolution}, authors have presented evolution analysis over five years of subreddits with respect to text, images, and links though they have only considered posts and not analyzed comments. Ref.~\cite{missing_data} has reported the effect of missing data and its implications over the Reddit corpus taken from 2005 to 2016.\\

\section{Data description}
\label{sec:data}

\subsection{Terminologies}
Following are the terminologies that are frequently used throughout the paper:
\begin{itemize}
	\item A Reddit \textbf{Post} can be text, link or a image submitted by a registered member. Posts are integral entities which allow users to express themselves and initiate a discussion. 
	
	\item \textbf{Comment} is a response to the post that is active on Reddit. A comment can either be a direct response to the post or a response to any comment made on a post, thus creating a nested structure of a tree graph.
	
	\item \textbf{Author} is a registered user on the platform who have at least one post or comment. 
	
	\item \textbf{Score} is the difference between number of upvotes and downvotes. 
    
        
\end{itemize}

\subsection{Definitions}
We define following quantities:
\begin{itemize}
	\item \textbf{Age} is the time difference between the last comment on the post and creation of the post, measured in seconds (unless otherwise mentioned).
	\item \textbf{Effective Comments} are the total number of comments on a post made by users other than the author of the post.
    \item \textbf{Automoderator} Official bot of Reddit
    \item \textbf{Deleted Author} Authors who delete their post/comment 
	
\end{itemize}

\subsection{Data}
We use two separate data sets of Reddit~\cite{Reddit_dump}, to see if the data shows any qualitative changes along with the quantitative changes, in a gap of few years -- 
\begin{itemize}
\item 
Period I: 1 January 2008 - 31 December 2008,
\item 
 Period II: 1 August 2014 - 31 July 2015,
\end{itemize}

The data contained posts and comments during those entire periods of one year each, and the associated variables like the title of the post, time of post/comment, subreddit, parent post/comment id, etc. Basic statistics of the data are presented in Table~\ref{tab:basic_data}. After September 2015, there was a change in the number of fields that were being provided by Reddit API. To maintain consistency, we have used the data till July 2015.

In this study, we have considered only those comments which were made on the posts of Period I during the same period and similarly for Period II. We also neglected the comments made during Period I to posts created before Period I and also comments made on the posts of Period I beyond the time domain of Period I. Same procedure was adopted for Period II. 
We analyzed the data using parallel computation on a Hadoop setup.


\begin{table}[h!]
	\begin{center}
		\caption{2008 Data Table}
		\label{tab:basic_data}
        \resizebox{\columnwidth}{!}{
		\begin{tabular}{|l|r|r|}
        \hline
            & Period I & Period II \\
			\hline
			Number of Posts & 2,523,761 & 63,118,764 \\
			\hline
			Posts with deleted authors & 425,770 (16.87$\%$) & 12,346,042 (19.56$\%$)\\
			\hline
			Posts with zero comments & 1,536,962 & 23,417,869\\
			\hline
			Posts with one comment & 591,489 & 9,011,332\\
			\hline
			Number of Comments & 7,242,871 & 613,385,507\\
			\hline
			Number of Comments on posts of the period & 7,224,539 & 608,654,680\\
			\hline
            Number of Disconnected Posts & 219 (0.009$\%$) & 1,380 (0.002$\%$)\\
            \hline
            Number of Removed Comments & 355 (0.004 $\%$) & 248,493 (0.04$\%$)\\
            \hline
		\end{tabular}
  }
	\end{center}
\end{table}

\begin{table}[h!]
	\begin{center}
		\caption{Used Data Variables for posts and comments}
		\label{tab:data_Variables}
		\begin{tabular}{|l|l|}
			\hline
			\textbf{Posts} & \textbf{Comments}\\
			\hline
			author & author\\
            created utc & created utc\\
             & link id\\
            name & name\\
            number of comments & parent id\\
			\hline
		\end{tabular}
	\end{center}
\end{table}

Table~\ref{tab:data_Variables} shows set of variables from the available metadata for comments and post that are used for our analysis. Available metadata contains 36 post variables and 21 comment variables. We have not considered score in our analyses, except for determining cyborgs.

\section{Analysis of aggregated data}
\label{sec:aggr}
\begin{figure}[t]
\centering	\includegraphics[width=0.91\linewidth]{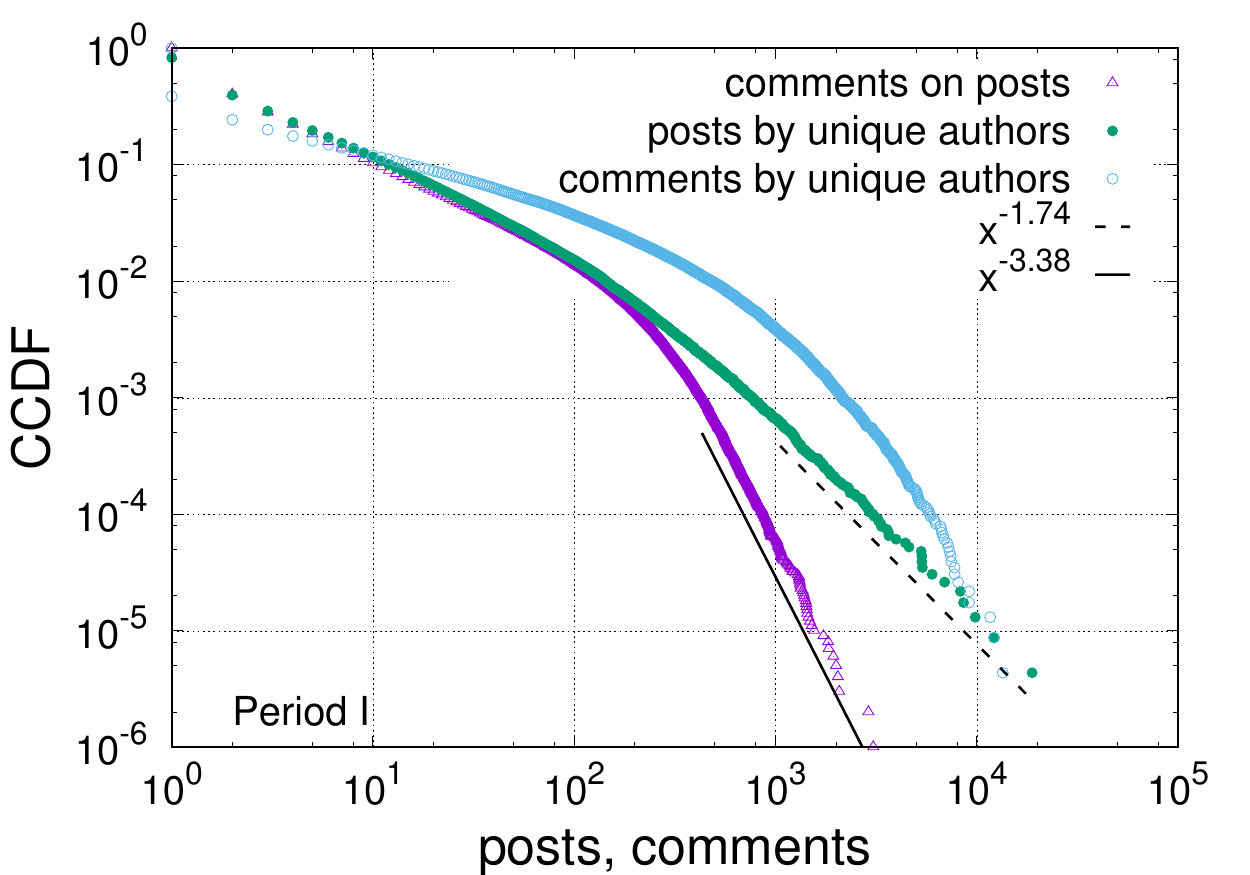}
\centering    \includegraphics[width=0.91\linewidth]{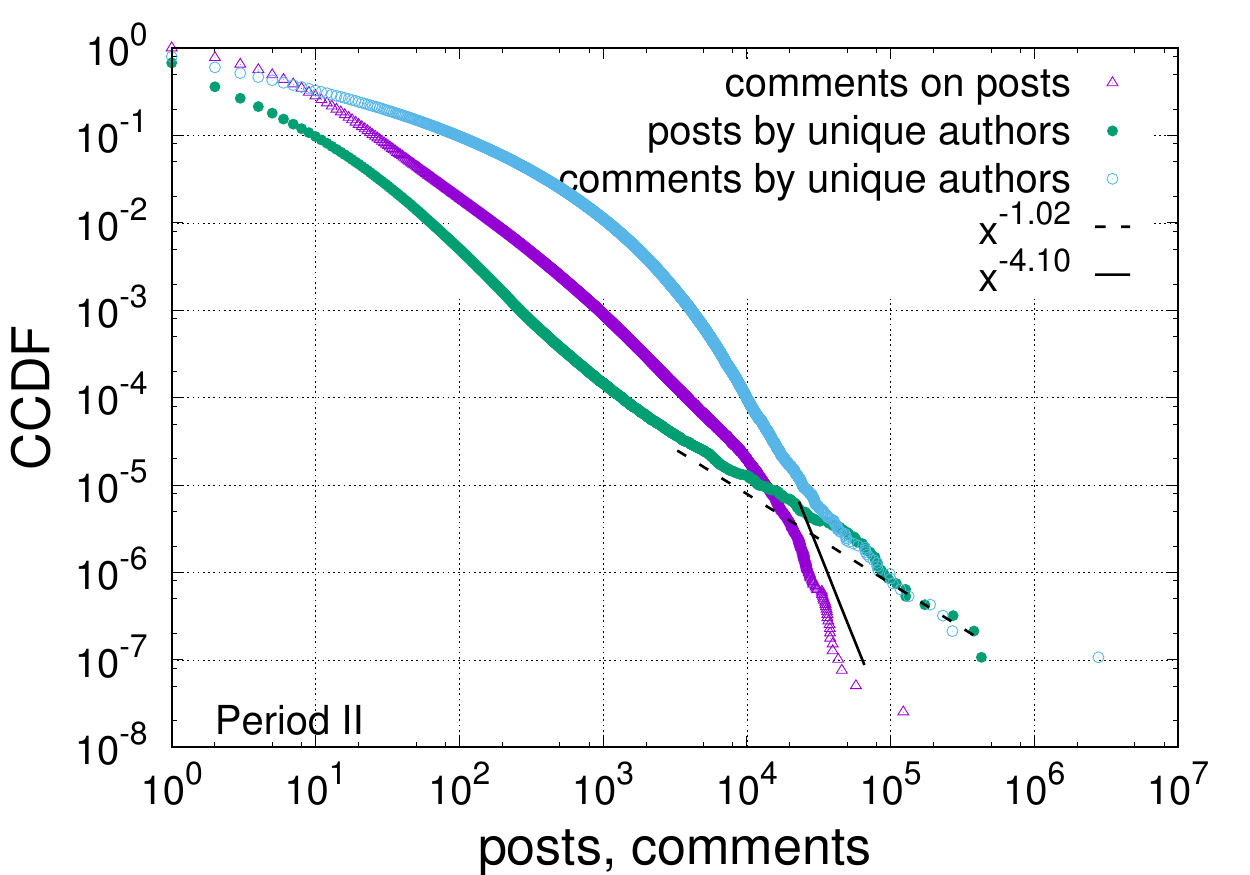}
	\caption{
		The basic distributions of posts, comments and authors. 
		Cumulative probability distribution (CCDF) that a post received at least $c$ comments, 
		CCDF that an author has posted at least $p$ posts, and 
		CCDF that an author commented at least $c$ times.
        Plots for both Period I and Period II are shown, along with estimates (using MLE) of fits to asymptotic power law tails.}
	\label{fig:ccdf_postcomment}
\end{figure}
For the analysis of one-year aggregated data, first we calculated the Complementary Cumulative Distribution Function (CCDF) which computes the probability that a post received at least $c$ comments. This is shown in Figure~\ref{fig:ccdf_postcomment}.  For Period I, with an average of $7.3$ comments per post, the CCDF shows a broad distribution with an asymptotic power law decay beyond $500$ comments: $Q(c) \sim c^{-\nu_c}$, with $\nu_c=3.38(1)$. 
For Period II,  with an average of $15.3$ comments per post, the tail of the CCDF also has a similar broad distribution with an asymptotic power law decay beyond $20,000$ comments: $Q(c) \sim c^{-\nu_c}$, with $\nu_c=4.10(3)$.
We can infer that while most posts get small number of comments, there are also significant yet diminishing number of posts  with large number of comments.
Apart from difference in the size of the data and the average number of comments per post, the asymptotic power law region commences much late in the Period II.


The probability (CCDF) that an author has posted at least $p$ posts also show an asymptotic power law decay for Period I: $Q(p) \sim p^{-\nu_1}$ with $\nu_1=1.74(2)$, with around $9.2$ posts per author on the average while for Period II, the asymptotic power law decay was $Q(p) \sim p^{-\nu_1}$ with $\nu_1=1.02(2)$, with around $6.7$ posts per author on the average. It is interesting to note a qualitative difference in the distributions -- while in Period I, the decay gets steeper after around $500$ posts, in Period II, it gets slower beyond $1000$ posts. 
However, the probability (CCDF) that an author commented at least $c$ times also shows broad distribution, but with a faster decay, resembling a lognormal (Figure~\ref{fig:ccdf_postcomment}) distribution for very high values, with around $21.2$ comments per author on the average for Period I, while for Period II the lognormal like behavior was followed by a slower decay with around $65.4$ comments per author on the average. This indicates a higher tendency to comment than to create a post. One of the reasons behind this is the less diversity of authors in contributing to posts compared to comments. Period II also shows the isolated data point for \textit{automoderator}.
The power law fits in our analysis are performed using \textit{maximum likelihood estimates} (MLE)~\cite{clauset2009power}.

While posting behavior is an intrinsic property of a user, and expected to be less correlated to comments, commenting is a part of the interaction with others and thus have a strong correlation with the behavior of others in the comment space, which justifies our observation that the former shows power law tail while the latter is lognormal. 


\begin{figure}[t]
\centering	\includegraphics[width=0.91\linewidth]{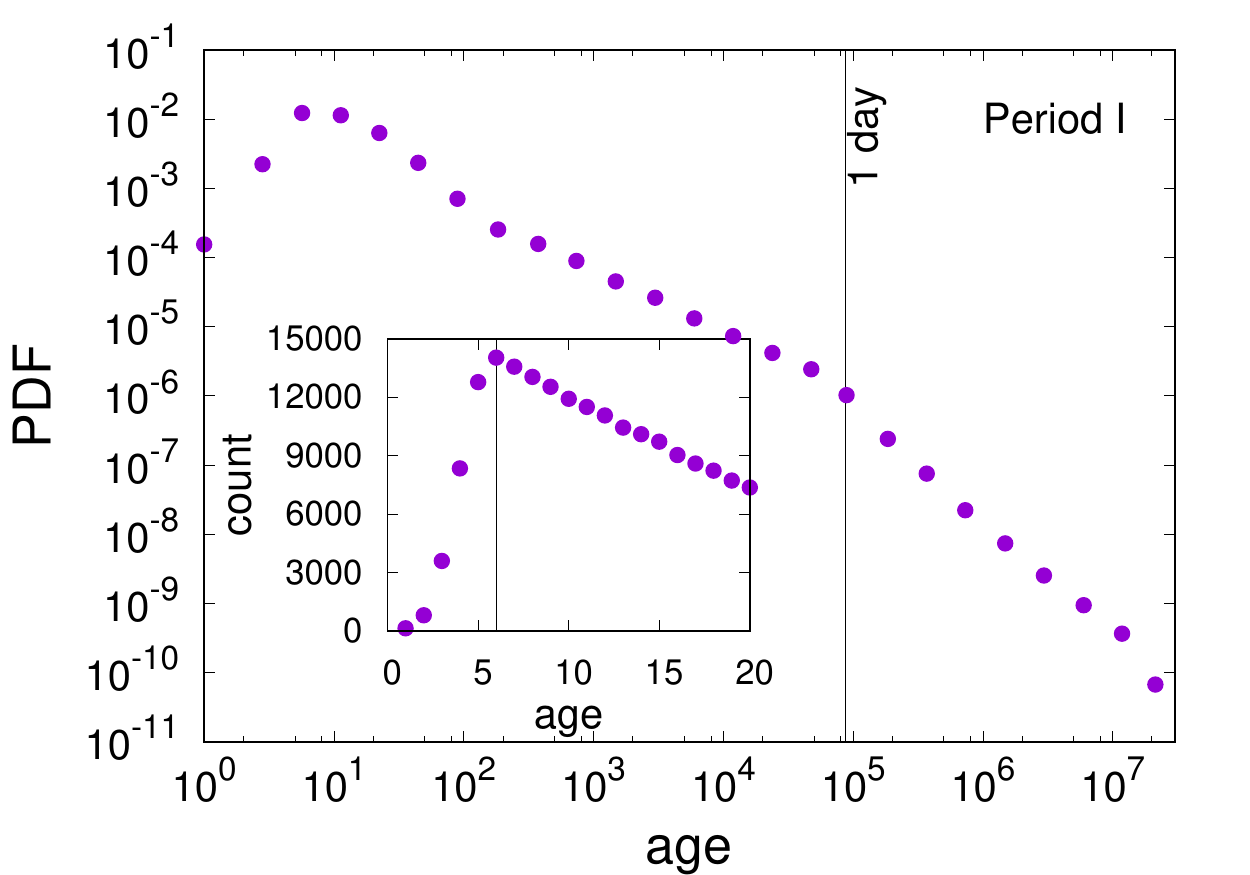}
 \centering   \includegraphics[width=0.91\linewidth]{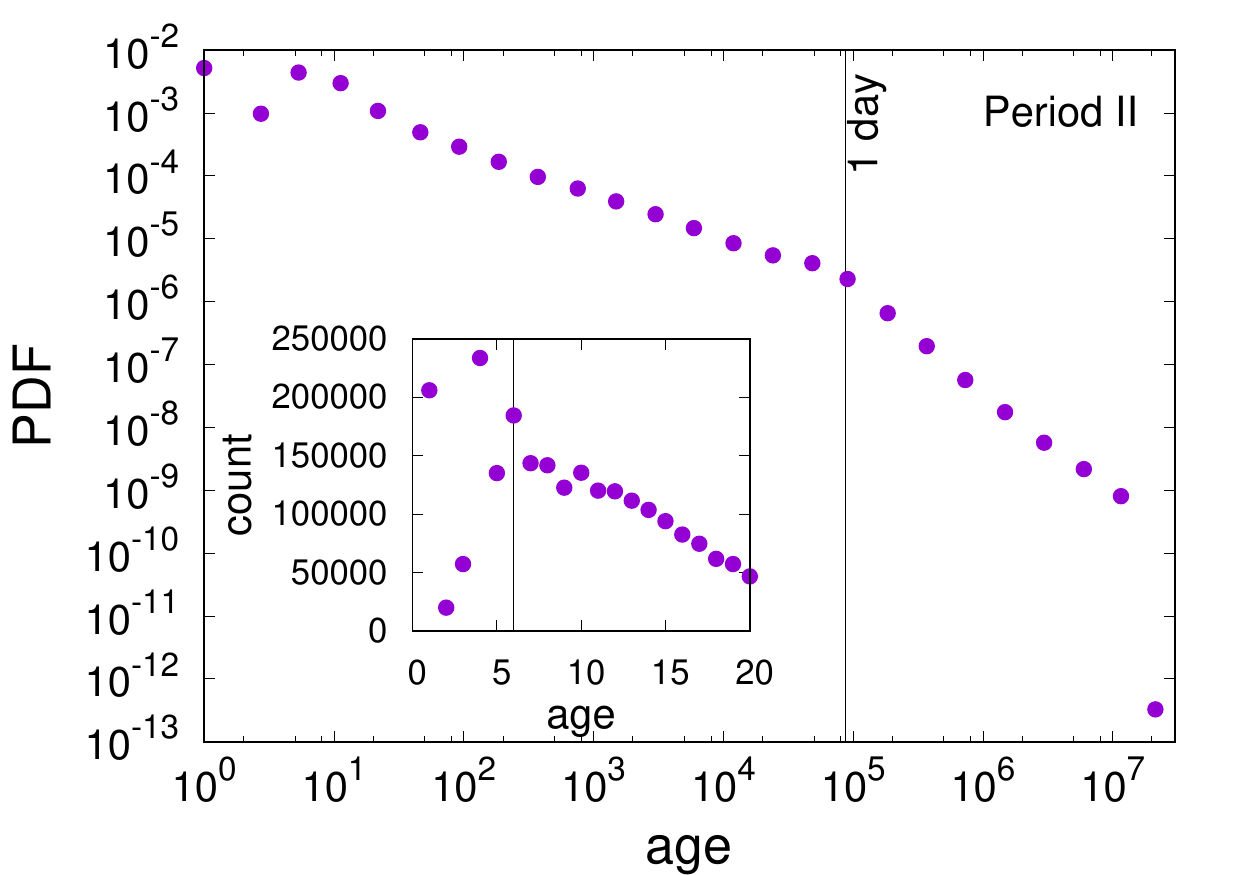}
	\caption{The main plots shows the PDF of the age of a post (seconds) for the entire age range for Periods I and II. 
		There is a marked departure in the nature of the probability distribution around $\approx 1$ day, indicating that a large number of posts become inactive beyond that time.
		The insets show the histograms corresponding to the age distribution at small values of age. There is a prominent peak around $6$ seconds for Period I and at values less than that for Period II. 
	}
	\label{fig:age_posts}
\end{figure}

\section{Analysis of post evolution patterns}
\label{sec:evolution_patterns}

To analyze the evolution of the posts, we calculate the age and number of comments for each post.


\subsection{Mayfly Buzz}
The probability density function (PDF) of the ages of all posts (Figure~\ref{fig:age_posts}) has a most probable value at $6$ seconds for Period I, while the equivalent peak is smeared across values less than $6$ seconds for Period II.
Also, we observe that there is also a shift in slope around age of 1 day, following which, the PDF decays faster, suggesting that more posts tend to become inactive after a day. In fact, $88.6\%$ in Period I and $71.1\%$ in Period II of posts die before a day. We term this behavior of the posts as \textit{Mayfly Buzz}, which resonates with the concept of creating a buzz for a day. Activity usually dies after a very short period of time, as seen in other social networking platforms. We observe the similar behavior on Reddit, where age is longer as we are dealing with a discussion platform as opposed to a microblogging site like Twitter~\cite{kwak2010twitter}, etc. 

\subsection{Cyborg-like behavior}



\begin{figure}[t]
\centering	\includegraphics[width = 0.95\linewidth]{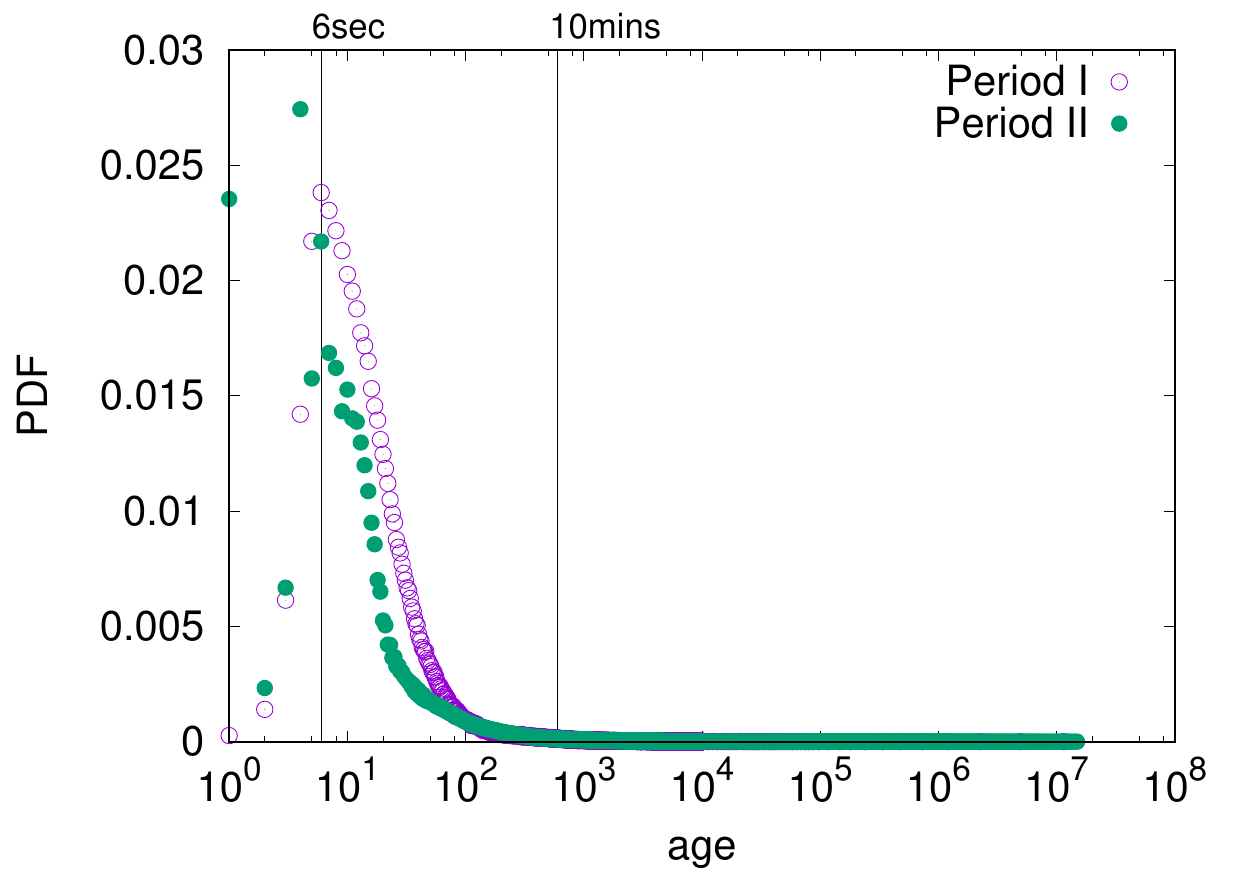}
	\caption{PDF of ages for posts with one comment for Periods I and II.}
	\label{fig:age_onecomment}
\end{figure}

\begin{table}[h]
	\begin{center}
		\caption{Cyborg-like Posts Statistics}
		\label{tab:Cyborg_data}
                \resizebox{\columnwidth}{!}{
		\begin{tabular}{|l|r|r| }
        	\hline
        	  & Period I  & Period II \\
			\hline
			Posts with first comment in less than 6 seconds & 43138 & 1,804,374\\
			\hline
			Posts with same author of first comment & 7,615 & 492,928\\
			\hline
			Cyborg-like Posts & 6,389 & 387,845\\
			\hline
			Successful Cyborg-like Posts &  3,446 & 70,237\\
			\hline
			Successful Non Cyborg-like Posts & 866 & 28,892\\
			\hline
			Unsuccessful Cyborg-like Posts & 2,943 & 317,608\\
			\hline
            Unsuccessful Non Cyborg-like Posts & 360 & 76,191\\
            \hline
		\end{tabular}
        }
	\end{center}
\end{table}
Figure~\ref{fig:age_onecomment} shows the age distribution (frequency) of all the posts which have only a single comment. In Period I, there is a very prominent peak at $6$ seconds, as found earlier (Figure~\ref{fig:age_posts}).
It can be seen that the ages of $72.78\%$ of these posts do not exceed $600$ seconds ($=10$ minutes). Period II looks very similar except the peak is seen at $5$ seconds with an additional peak at $1$ second.

Further, we analyzed posts whose first comment is posted within $6$ seconds, which constitutes $43138$ posts for Period I and $1,804,374$ for Period II. Out of these posts we found that there are $7,615$ and $492,928$ posts which have their first comment by the author of the post for Period I and Period II respectively. We observe an uncanny behavior from approximately $17\%$ and $20\%$ of people behaving in exactly the same manner respectively. To understand this uncanny behavior of posting comment by the same user, we checked the number of characters in the first comment of these posts. For instance, we find that $83.9\%$ ($6,389$ of $7,615$) and $79\%$ ($387,845$ of $492,928$) posts have number of characters more than $100$ for Period I and II respectively. Writing such long comments within $6$ seconds is quite impossible for a genuine human.  We categorize these posts to be exhibiting a \textit{cyborg-like} behavior, where these posts may be just an advertisement or a message that these users need to propagate.

\begin{figure}[h]
\centering	\includegraphics[width = 0.90\linewidth]{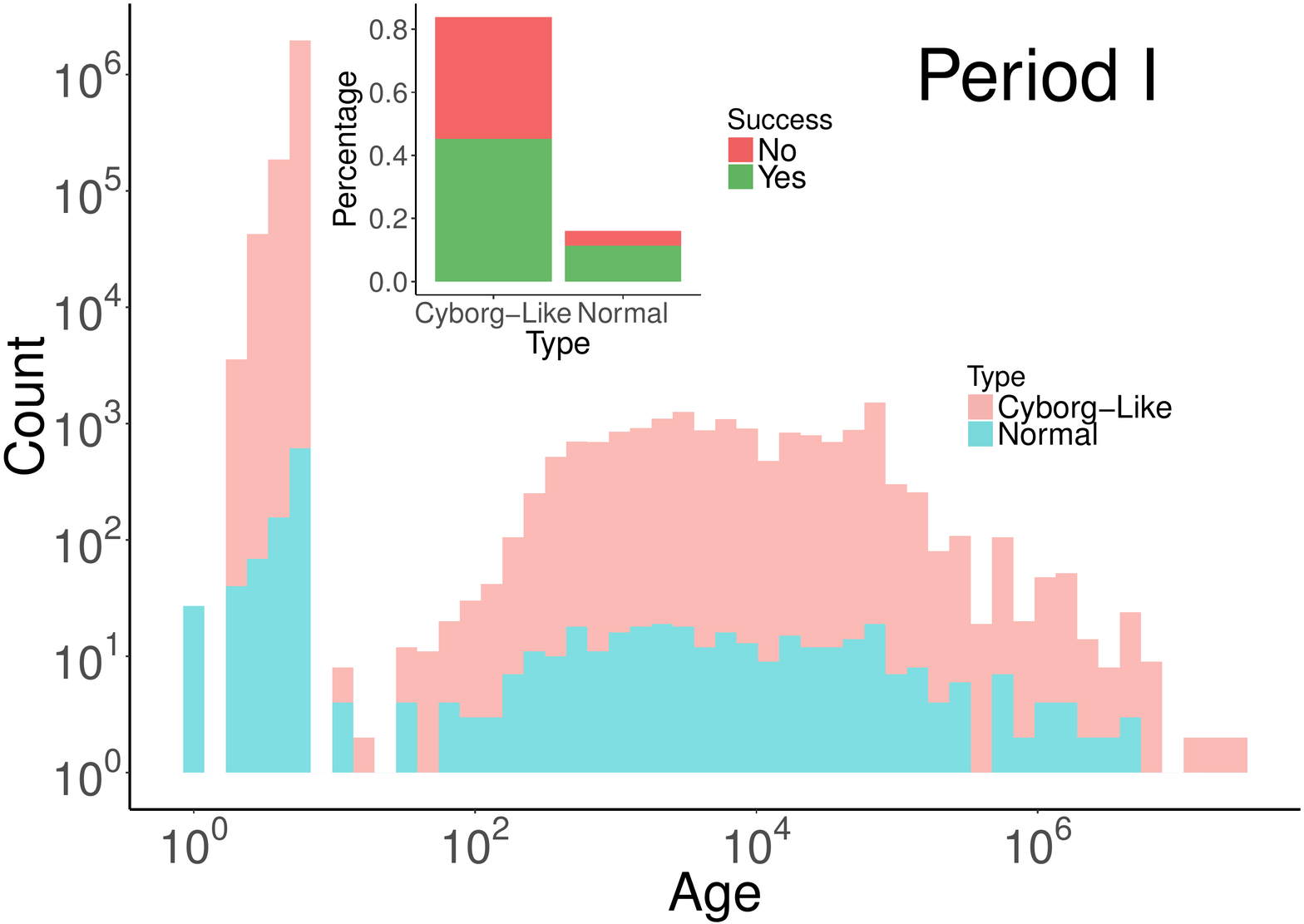}
\centering     \includegraphics[width = 0.90\linewidth]{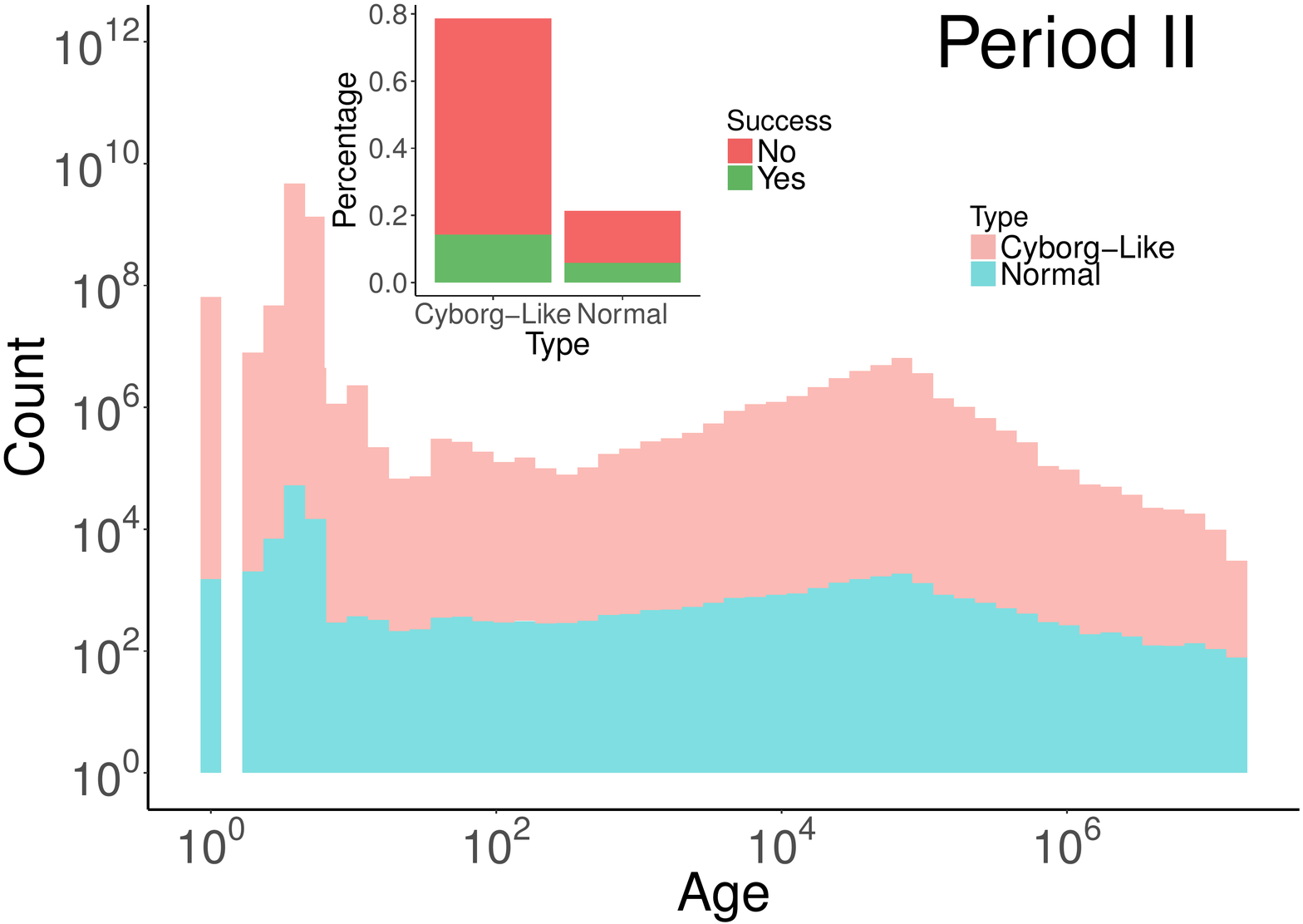}
	\caption{The histograms of actual age of the posts whose comment is within $6$ seconds. The insets show distribution of posts over success rate. Plots are shown for both Period I and Period II.}
	\label{fig:bot_inset}
\end{figure}
We further define a success criterion to check if these posts where successful in garnering attention or responses, if a post is getting any reaction (comment or vote) from other Reddit users, then they are considered successful in drawing attention. For instance, we find that $53.93\%$ ($3,446$ of $6,389$) and $18\%$ ($70,237$ of $387,845$) \textit{cyborg-like} posts were  successful in Periods I and II respectively. While $70.63\%$ ($866$ of $1,226$) of normal posts (which have comments with less than 100 characters) of Period I are successful, which we assume can be possibly done by humans (Figure~\ref{fig:bot_inset}), which comes to $27.5\%$ ($28,892$ of $105,083$) for Period II. Table~\ref{tab:Cyborg_data} summarizes the data for this analysis. 
Hence, for Period I, we infer that machine generated content is less likely to garner interest as compared to human generated content. A possible reason behind the low success of the cyborg-like posts can be that lengthy comments and promotions/advertisements provide less room for any discussions. 
For Period II, however, we found a variety of behavior in the cyborg-like posts in the posts, which required much more granular analysis. 

\begin{figure}[h]
\centering	\includegraphics[width = 0.95\linewidth]{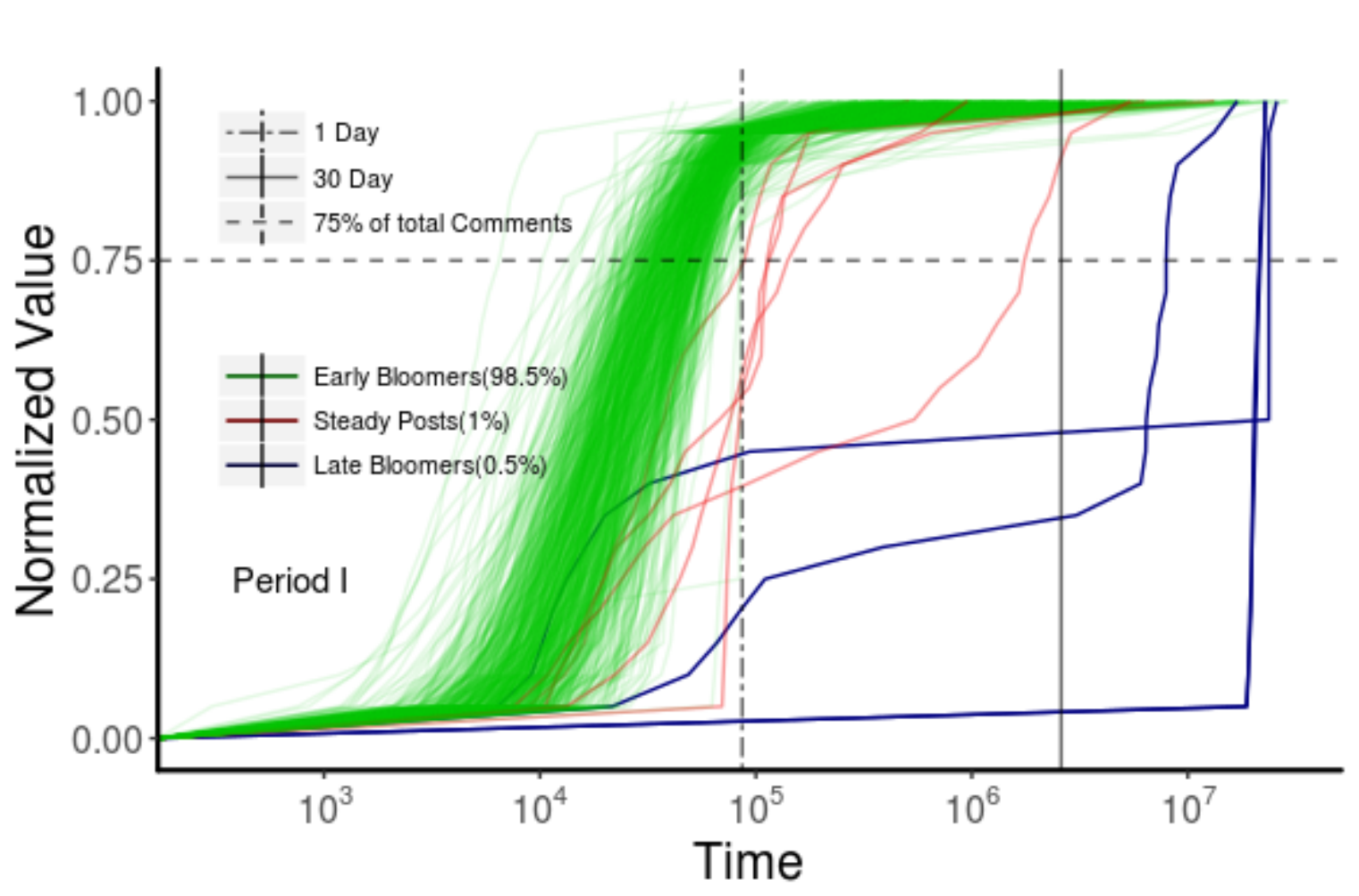}
\centering    \includegraphics[width = 0.95\linewidth]{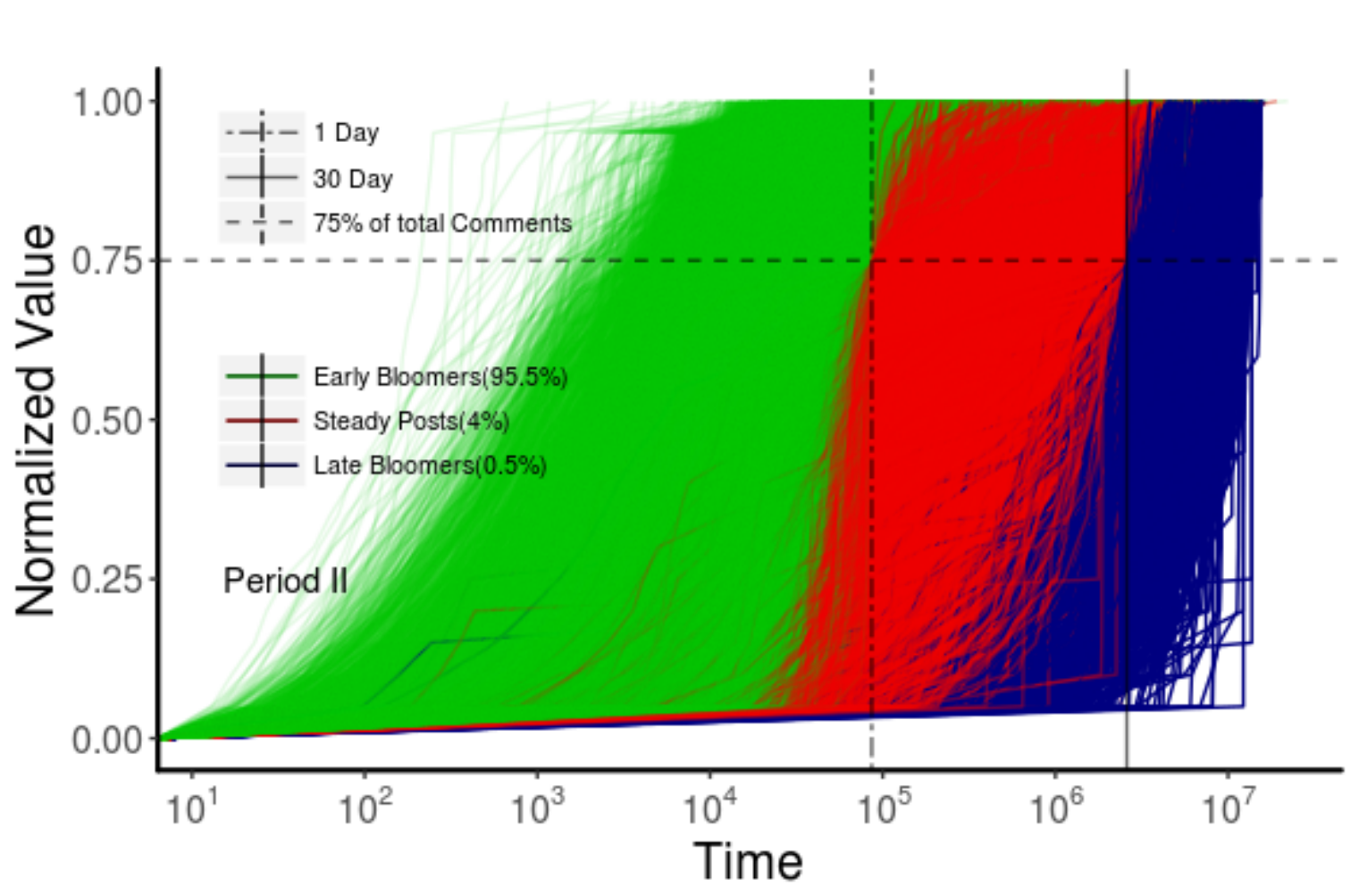}
	
	\caption{
		Time evolution of the number of comments in a post (normalized by the final number of comments obtained in our defined time window of 1 year) for posts with more than $500$ comments. The data has been coarse-grained to simplify visualization. The horizontal line corresponds to 75\% of the total comments and the vertical lines are at 1 day and 30 days. Some posts are mostly active within 1 day (green), some  grow throughout their active life span (red), while others grow slowly while becoming active at some later stage (blue). Plots are shown for Period I (top) and Period II (bottom).
        }
	\label{fig:comments_age_time1}
\end{figure}
        
\begin{figure}
\centering \includegraphics[width = 0.95\linewidth]{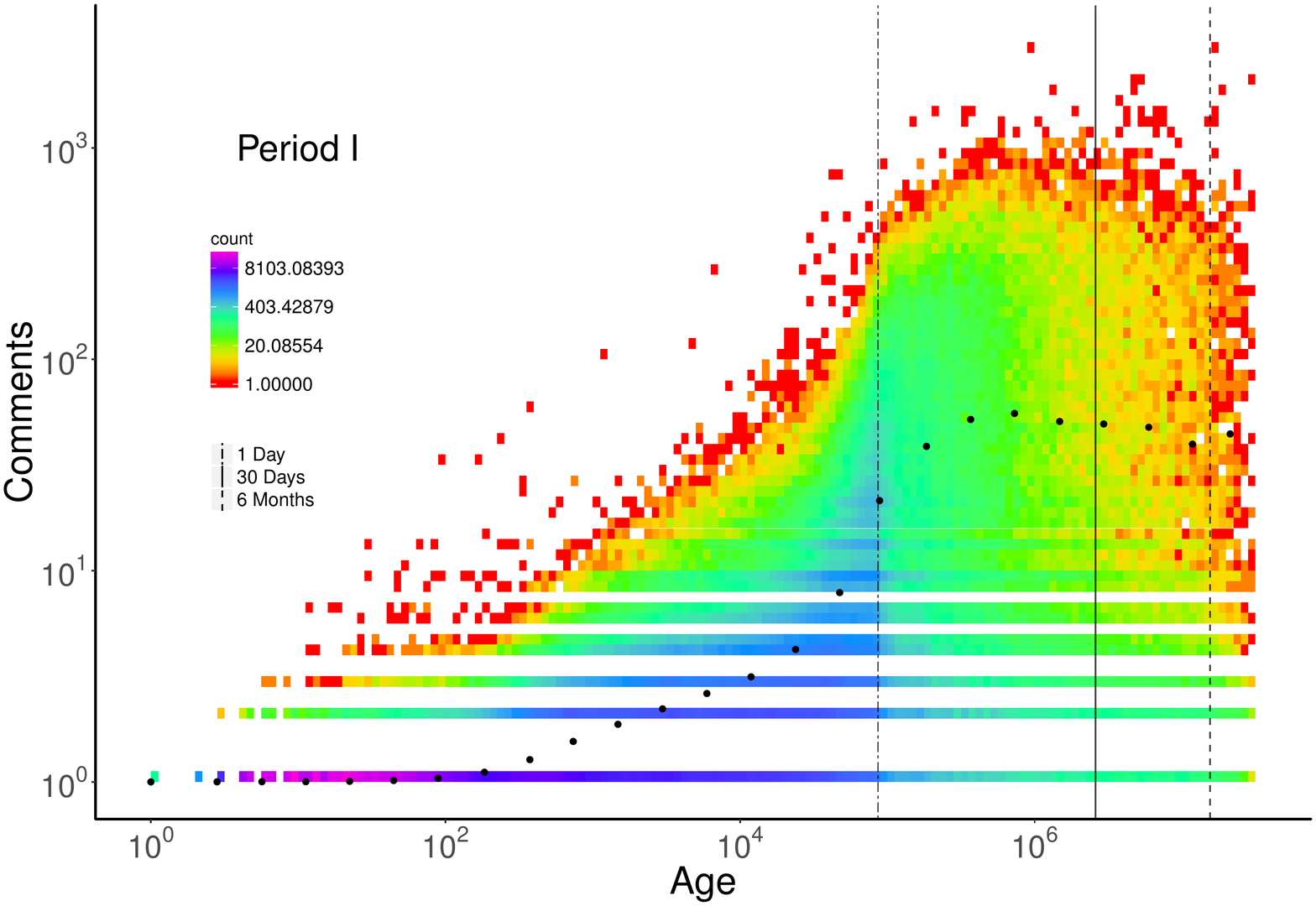}
\centering \includegraphics[width = 0.95\linewidth]{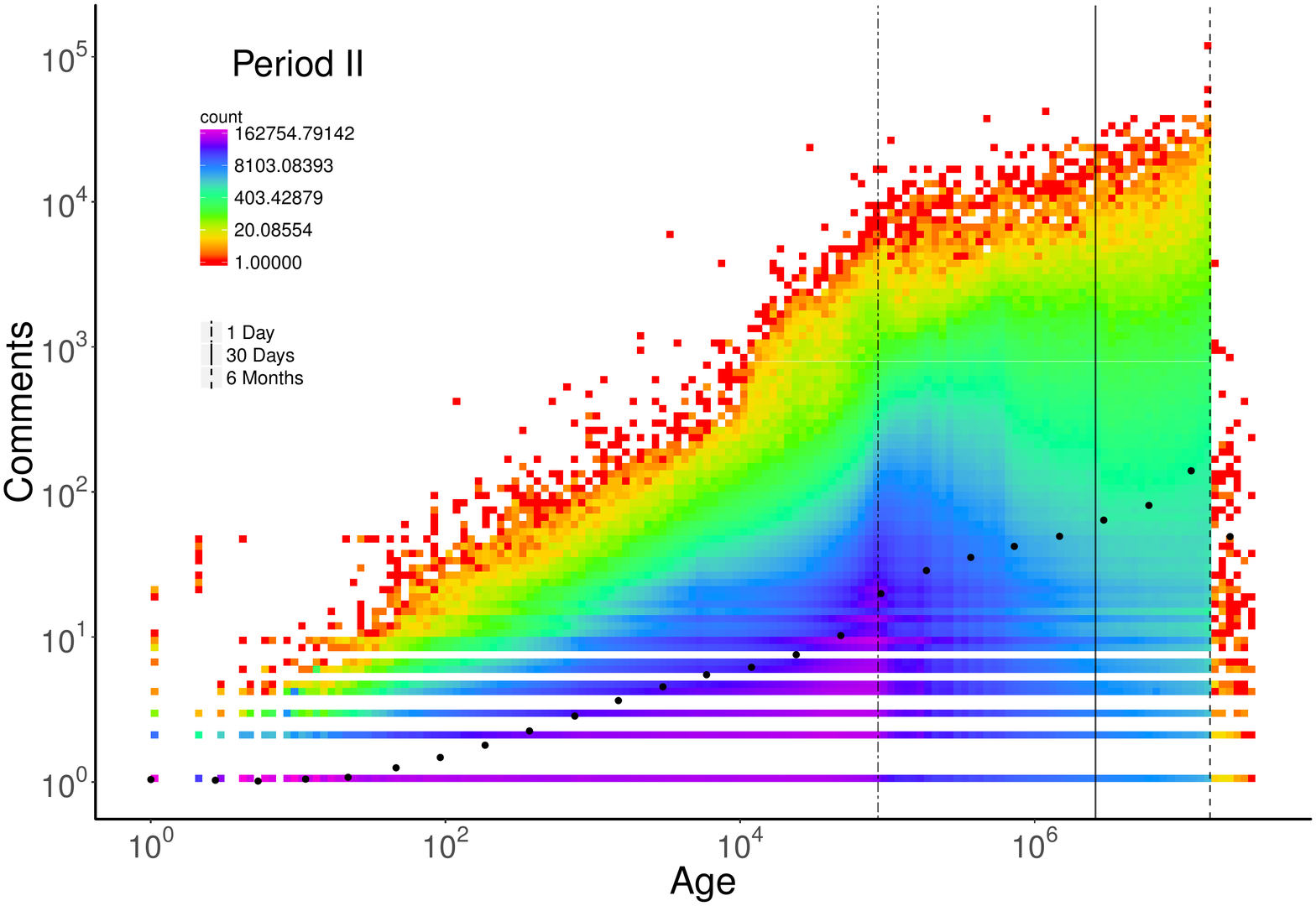}
		\caption{The number of comments and the age of a post is shown as a density heat map for all posts, along with its binned average. A marked departure in the gross behavior around 1 day is prominent in both Period I and Period II.
	}
	\label{fig:comments_age_time2}
\end{figure}

\subsection{Popular Post Dynamics}

To understand the age dynamics of the popular posts and infer their behavior, we have plotted the time evolution of the posts which have more than $500$ comments. We observe that there are three distinct categories (Figure~\ref{fig:comments_age_time1}):
\begin{itemize}
	\item \textit{early bloomers} are rapidly growing posts, accumulating more than $75\%$ of their total comments within $1$ day, creating the \textit{Mayfly Buzz} as discussed earlier,
	\item \textit{steady posts} are characterized by activity throughout their lifespan.
	\item slowly growing posts, which get suddenly very active at a late stage (30 days), can be termed as \textit{late bloomers}. 
\end{itemize}
We also study the behavior of the total number of comments with the age of each post, for all posts in our data. Figure~\ref{fig:comments_age_time2} shows the heat map for all posts. The overlaid binned average of all data indicates a marked departure in the gross behavior around 1 day which is also prominent from the density in the heat map.


\section{Analysis of interactions}
\label{sec:interaction}



The Reddit post-comment structure forms a tree graph, where posts can have its comments, and the comments can further garner replies. For this analysis, we have calculated a \textit{limelight score} for each post based on the number of comments gathered as reply to a single first-level comment. In a way, this score computes the depth of discussion around a single comment for a post. 

\begin{equation*}
\textrm{Limelight Score} = \frac{\max(Comm_j)}{\sum_{k = 1}^{N} Comm_j}
\end{equation*}
where $Comm_j$ is the total number of comments under $j^{th}$ first level comment and $N$ is the total number of first level comments for that post.


\begin{figure}
 \centering   \includegraphics[width = 0.90\linewidth]{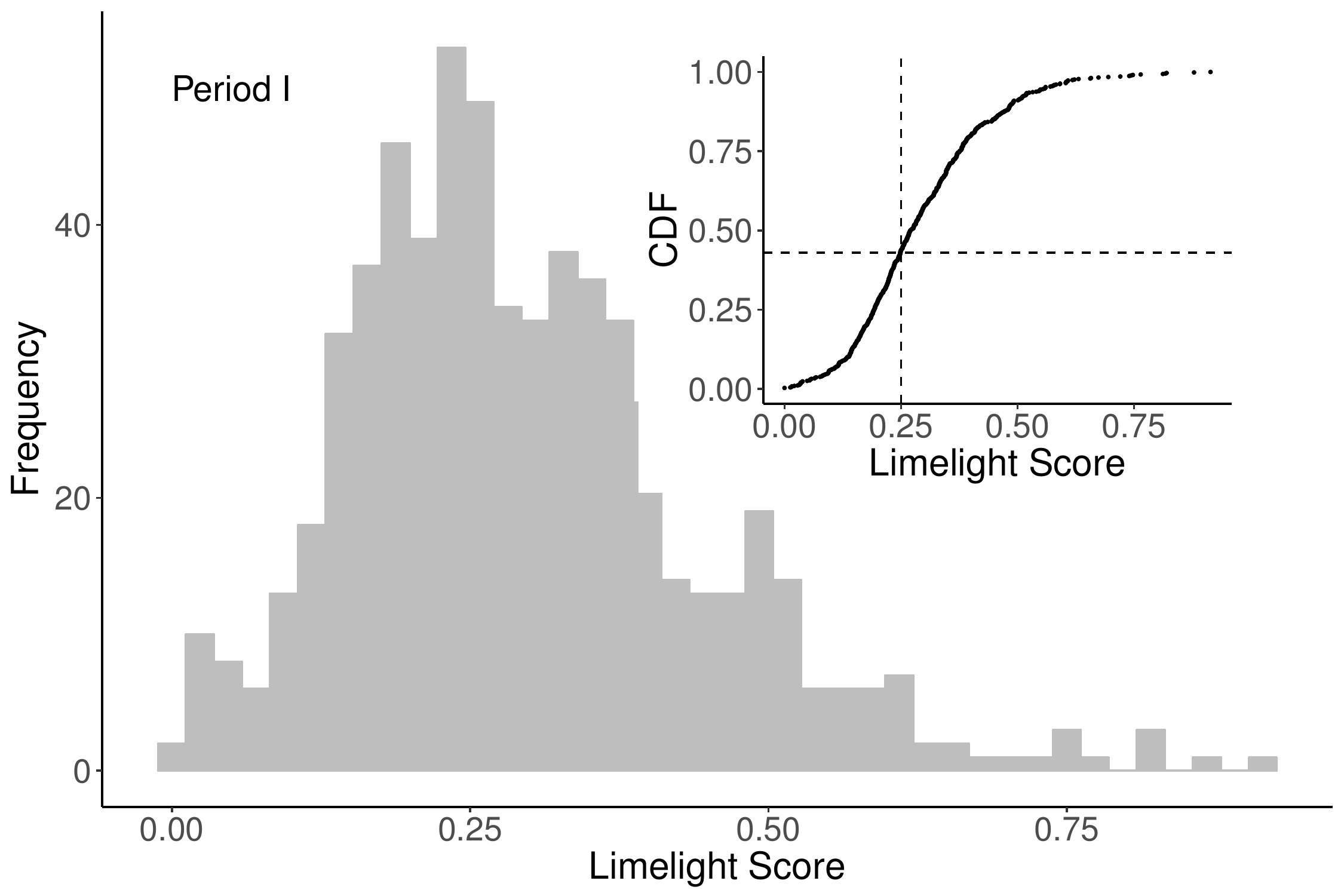}
 \centering   \includegraphics[width = 0.90\linewidth]{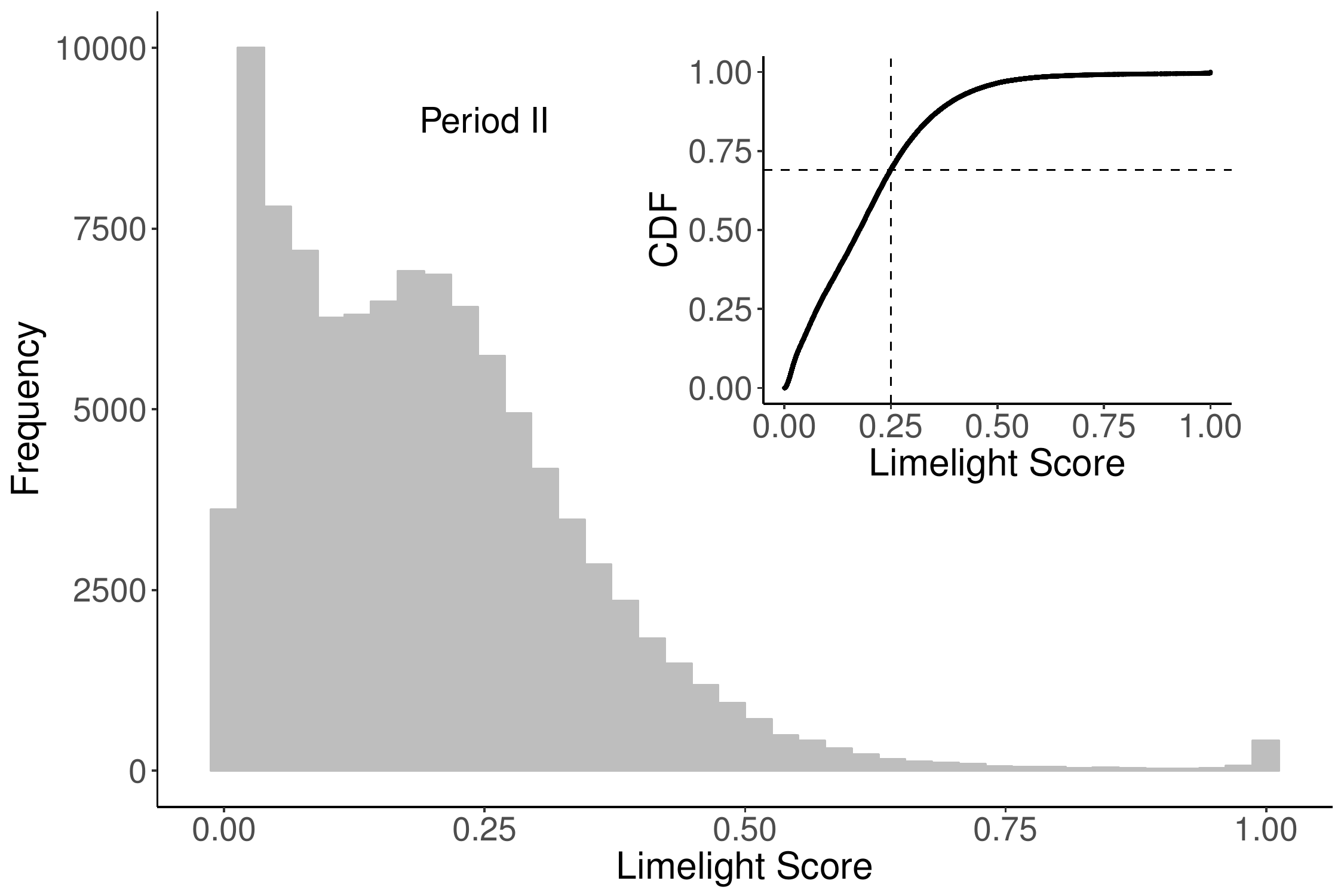}
    \caption{Histograms of Limelight scores for the $628$ posts for Period I and $100,422$ posts in Period II, which have more than $500$ comments. The insets show the corresponding CDFs.}
    \label{fig:limelight}
\end{figure}
Figure~\ref{fig:limelight} shows the histogram of the \textit{Limelight scores} while the inset shows the CDF of the same. Here we have considered posts that have 500 or more comments only. 
We observe that in Period I, $56\%$ of the total posts contain one comment with \textit{Limelight score} of at least $0.25$, which means that at least 25\% of the discussion in this post is initiated and centered around a single comment.
This behavior is also exhibited in $31\%$ of the posts in Period II.  Additionally in Period II, a finite number of posts actually have Limelight score close to unity, indicating absolute dominance of one branch of the comment tree.

We also observe that most of the time, the author of first level \textit{Limelight} hogging comment is not the author of the post. For instance, this is true for about $97\%$ of the posts during Period I. 

This leads to an interesting insight that links virtual human behavior in social media to physical world social behavior. It is a rather common scenario that during any group discussion or meeting usually there are a few specific people, other than the presenter, who pro-actively initiates a conversation asking a question or making a comment, whereafter other people join the conversation. Interestingly, it is observed that lime-light hogging behavior is completely missing for posts whose authors exhibit \textit{Cyborg-like} behavior. 
Thus, it may be inferred that posts automatically generated by bots have failed to garner garner human attention most of the times. However, we will conduct more rigorous studies in future to validate the inference.

To the best of our knowledge, characterizing content popularity by the depth of discussion around it has not been attempted before. Since it has been proved in earlier studies~\cite{stoddard2015popularity,glenski2017consumers} that the number of upvotes-downvotes are not meaningful indicators for measuring interestingness or popularity of content, we claim that this can be a good way to measure them. 



\section{Analysis of Author Behavior}
\label{sec:author}
For analyzing author interactions, we define a network where, nodes represent unique authors and the edges represent the interaction between the authors through comments. We define the in-degree and out-degree for each node based on the number of interactions, where a self loop is ignored. Table~\ref{tab:author_data} shows the statistics for the $3$ categories -- (i) authors who only put up posts are the pure \textit{content producers}, (ii) authors who only comment are the pure \textit{content consumers}, and (iii) rest of them indulge in both of the activities. 


%
\begin{table}[h!]
	\begin{center}
		\caption{Author Table}
		\label{tab:author_data}
                \resizebox{\columnwidth}{!}{
		\begin{tabular}{|l|r|r|}
        	\hline
        	  & Period I & Period II \\
			\hline
			Total Active Authors & 229,488 & 9,369,708\\
			\hline
			Total Authors who only create posts & 140,918 & 1,917,161\\
			\hline
			Total Authors who only comments & 39,764 & 3,019,676\\
			\hline
			Total Authors who comment as well create posts & 48,806 & 4,432,871\\
			\hline
		\end{tabular}
        }
	\end{center}
\end{table}


\subsection{Quantifying author interactions to assess their influence}

\begin{figure}[h]
\centering	\includegraphics[width = 0.93\linewidth]{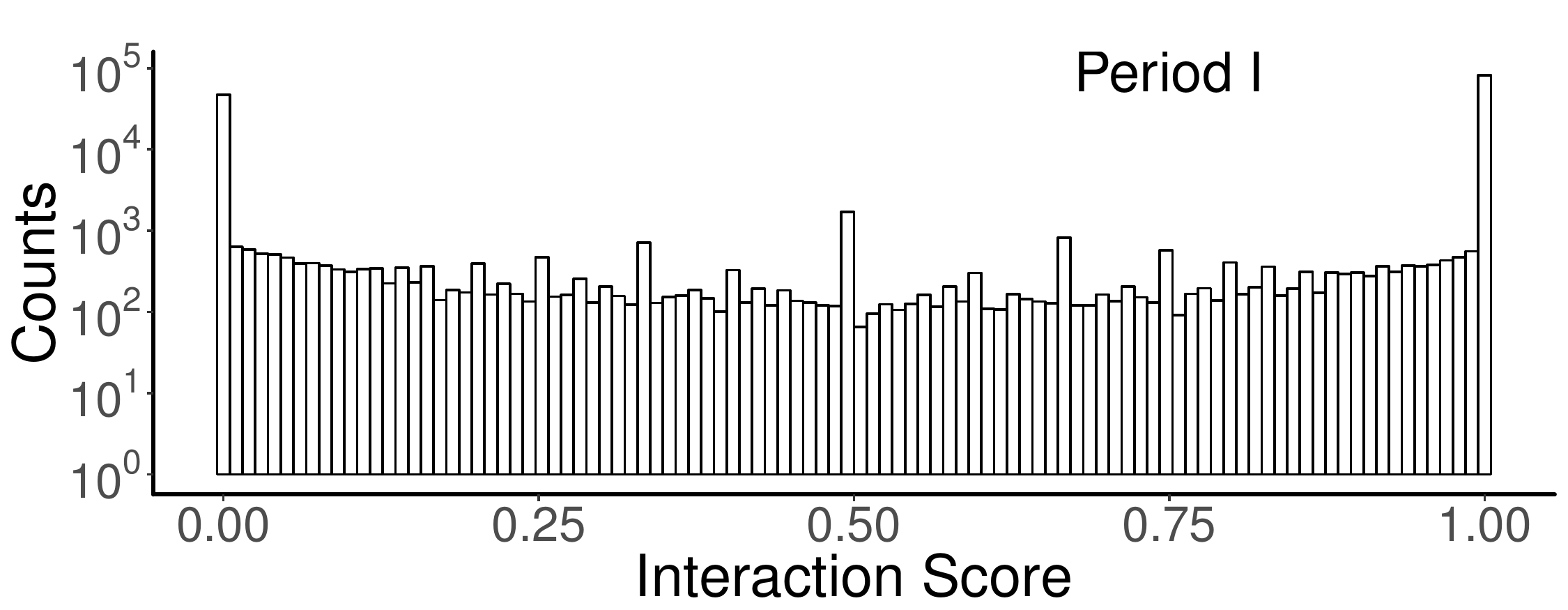}
\centering    \includegraphics[width = 0.93\linewidth]{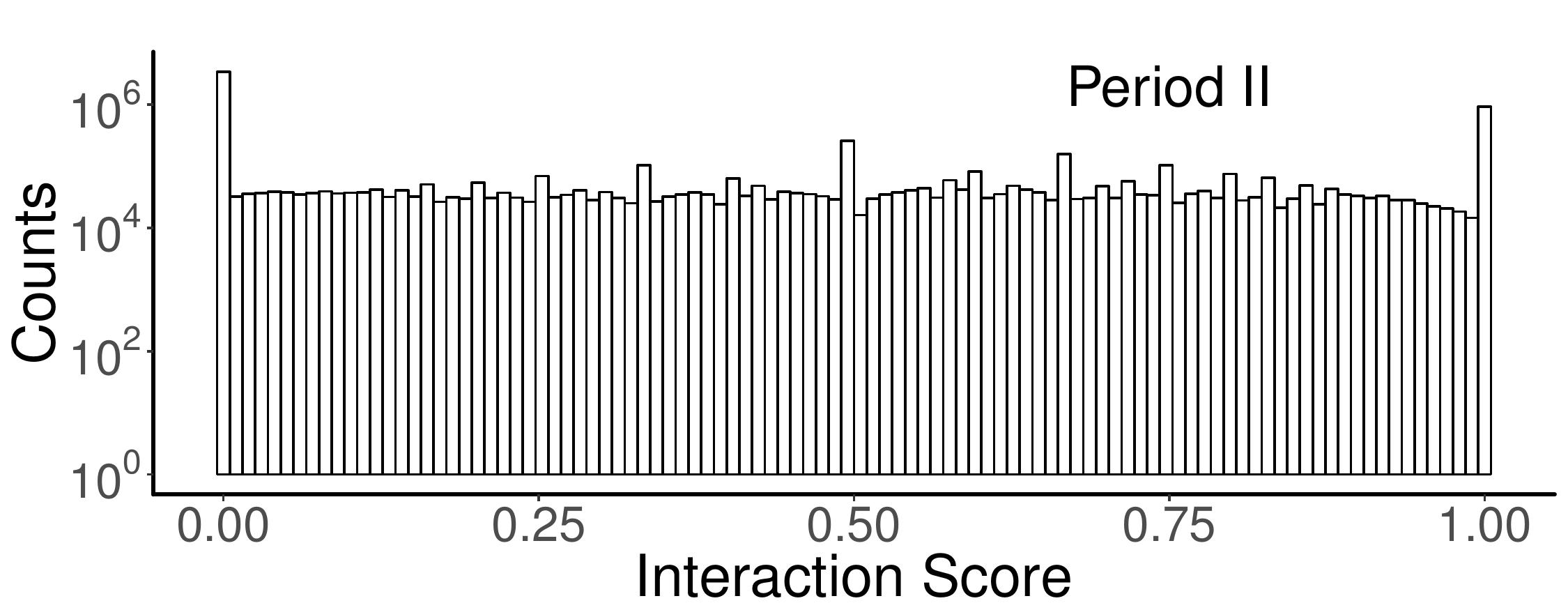}
	\caption{Frequency counts for different interaction scores of authors for both periods. Equal reciprocative behavior is observed at score $0.5$.}
	\label{fig:interaction_score}
\end{figure}

If $A=$ total effective number of comments received and $B$ = total number of comments on others' posts, then we define the \textbf{interaction score} of an author as $A/(A+B)$. Interaction score is zero for all authors who comment on others' posts but have not received any comments on their posts. Score is $1$, if an author does not comment on others' posts but receives comments on one's own posts, though this is rarely observed. Figure~\ref{fig:interaction_score} shows the histogram for the total count of authors along the whole range of interaction score.  

\begin{figure}
	\includegraphics[width = \linewidth]{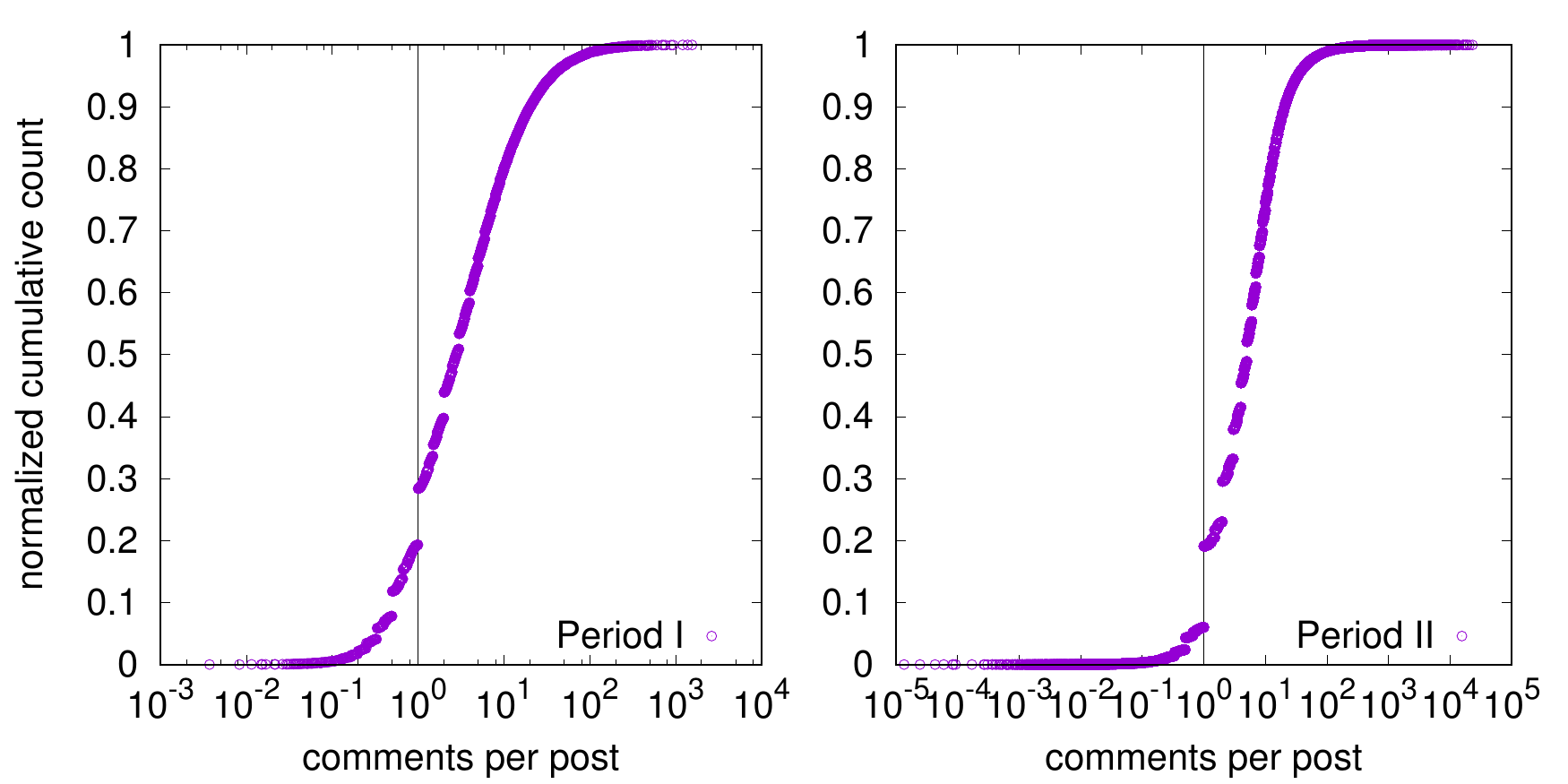}
	\caption{Cumulative ratio of authors according to comments per posts. The vertical line corresponds to unity. 
    For Period I, $22\%$ fall below unity, $11\%$ equals unity and the rest $67\%$ are over unity.
    For Period II, $6\%$ fall below unity, $13\%$ equals unity and the rest $81\%$ are over unity.
    }
	\label{fig:cummulative_ratio}
\end{figure}
There are some distinct authors who have the ability to consistently garner a large number of comments on each of their posts. To quantify this, we analyze the average number of effective comments received per post by authors. Figure~\ref{fig:cummulative_ratio} shows the normalized cumulative count across the effective number of comments per post. It is observed that $22\%$ of the authors have fewer effective comments than the number of posts that they have put up which means no interaction for many posts for Period I, which happens to be around $6\%$ in Period II. $11\%$ have received an equal number of effective comments as their posts which can be attributed to an average of one comment per post for Period I, which is $13\%$ for Period II. The rest $67\%$ received more comments than the number of posts put up for Period I, which comes to $81\%$ for Period II. 

The discussion above shows that authors who receive more attention on their posts are also the ones who are commenting on others posts. In other words, to gain attention on social media, authors have to be reciprocative. This is also highlighted by the peak at interaction score  of $0.5$ shown in Figure~\ref{fig:interaction_score}. It also emphasizes that the social media interactions are dominated by the phenomenon of mutual gratification.

\section{Conclusions}
\label{sec:conclusion}

Reddit, the large, community-driven social network and discussion platform, harbors a plethora of behaviors as far as users are concerned. While a huge fraction of posts are left uncommented, the distribution of the number of comments on posts show correlation through the power law tail. Behavior of authors show a large variety -- while many authors simultaneously post and comment, there are also a large fraction of purely \textit{content producers} and \textit{content consumers}, who restrict themselves only to posting and commenting respectively.
The authors show a strong correlation between themselves and indication of the underlying multiplicative process in the form of lognormal distribution for the largest values for the distribution of number of comments by unique authors. 
Each post stay active as comments flow in and discussions are produced. However, a huge fraction of posts have only a single comment, and among them, a majority receive that only comment within $6$ seconds, indicating a \textit{Cyborg-like} behavior. A large fraction of posts seem to become inactive around the age of $1$ day. This is consistent with the average active time of posts reported for micro-blogging site such as Twitter~\cite{kwak2010twitter}. When we look at the time evolution of the top commented posts, we find three broad classes of posts -- (i) \textit{early bloomers} who gather more than $75\%$ of their lifetime comments within a day, (ii) \textit{steady posts} growing steadily throughout their lifespan, and (iii) \textit{late bloomers} who show very little activity until the end of their lifespan. The early bloomers contribute to what we term as \textit{Mayfly Buzz}, and constitute the majority of the posts.
Posts also show \textit{limelight hogging} behavior and upon appropriate 
characterization, we find that $56\%$ for Period I and $31\%$ for Period II of posts have \textit{limelight score} above $0.25$, indicating that in such a large fraction of posts, at least one-fourth of the total weight of the discussions are contributed by one of the first level comments.
In fact, this measure can be a more meaningful indicator of interestingness or popularity of the content, compared to votes or only comments.


With the increasing use of social media even within closed groups as well as organizations, understanding human behaviors and able to characterize them is turning out to be an important task with potential impact and applications. One possible application of understanding temporal patterns of group behavior in such a scenario can be focused on injecting the right content or advertisement for the right group at the right time.

Our rigorous statistical analysis brings out a variety of behavioral elements from the authors and their interactions. There are few authors who are able to generate quite a lot of activity across a large number of posts. Going ahead, trend analysis of changing sentiment can be interesting. The insights gained from this analysis can be used to model different aspects from a large interactive population. In addition, predicting the recent trends can lead to better targeted reach e.g., innovative usage of \textit{memes}.



\end{document}